\documentclass[12pt]{article}
\usepackage{fleqn,amssymb,amsfonts,latexsym}
\usepackage{color}

\newenvironment{assert}{\begin{itemize}\item[]\it}{\end{itemize}}

\newcommand {\bR}{{\Bbb R}}
\newcommand {\bN}{{\Bbb N}}
\newcommand {\bZ}{{\Bbb Z}}
\newcommand {\bI}{{\Bbb I}}

\newcommand {\bP}{{\Bbb P}}

\newcommand {\bU}{{\Bbb U}}
\newcommand {\bV}{{\Bbb V}}

\newcommand {\cB}{{\cal B}}

\newcommand {\cF}{{\cal F}}

\newcommand{\beq}{\begin{equation}}
\newcommand{\eeq}{\end{equation}}
\newcommand{\beqn}{\begin{eqnarray}}
\newcommand{\eeqn}{\end{eqnarray}}
\newcommand{\beqno}{\begin{eqnarray*}}
\newcommand{\eeqno}{\end{eqnarray*}}
\newtheorem{theorem}{Theorem} [section]
\newtheorem{lemma}[theorem]{Lemma}

\newtheorem {remark}[theorem]{Remark}

\begin{document}
\title {On the Dynamics of Crystal Electrons,\\
        high Momentum Regime}
\author{Joachim Asch${}^\ast$\and Fran\c cois Bentosela${}^\ast$\and
Pierre Duclos\thanks{CPT-CNRS, Luminy Case 907,  F-13288 Marseille
Cedex 9, France. e-mail: user@cpt.univ-mrs.fr}
\and  Gheorghe Nenciu
\thanks{{Dept.Theor.Phys. University of Bucharest,
P.O.Box MG 11, 76900- Bucharest, Romania. e-mail: nenciu@barutu.fizica.unibuc.ro}} }
\date{26.1.2000}
\maketitle
\begin{abstract}We study the quantum dynamics generated by
$H^{SW}=-{d^2\over dx^2}+V-x$ with  $V$ a real periodic
function of weak regularity. We prove that the continuous spectrum of
$H^{SW}$ is never empty, and furthermore that for $V$ small enough
there are no bound states. 
 \end{abstract} %

 \section{Introduction}
Consider for $\alpha\ge0$   potentials of the form
\begin{assert}
$V:\bR\to\bR, V(x+\gamma)=V(x)\quad(x\in\bR,\gamma\in2\pi\bZ)$
with
$$\Vert V\Vert_\alpha^2:=\sum_{n\in\bZ}\vert\widehat
V(n)\vert^2(1+n^2)^\alpha<\infty$$
 where  $\widehat V(n):={1\over \sqrt{2\pi}}\int_0^{2\pi}e^{-inx}
V(x)\ dx$.
\end{assert}
The Stark Wannier Hamiltonian is the
selfadjoint operator
$$H^{SW}:=-{d^2\over dx^2}-x+V$$
defined in $L^2(\bR)$ by extension from the core
$C_0^\infty(\bR)$; this is a corollary of the Faris Lavine
theorem, see \cite{reedsimo2}.

We shall prove that there are always propagating states and that 
small potentials with some regularity
cannot bind:
\begin{theorem}\label{cont}
\begin{itemize}
\item[(i)] Let $\alpha>0$ then
\[\sigma_{cont}(H^{SW})\neq\emptyset;\]
\item[(ii)] for $\alpha>1/2$ there is a $c>0$ such that for ${\Vert
V
\Vert_\alpha}<c$ 
\[\sigma_{pp}(H^{SW})=\emptyset.\]
\end{itemize}

\end{theorem}

This theorem is a consequence of our main result, Theorem
\ref{main}, which asserts that the probability to be
accelerated in the future grows with the momentum of the
initial state.

Remark that our results are stated  in terms of the Stark--Wannier
problem  but apply as well to the problem of driven quantum rings,
see \cite{avronemi}.

The dynamics of crystal electrons described by the present model
have been studied since \cite{wannier} both in mathematics and
physics literature, see \cite{nenc91} for a review. The general
problem is to  understand how the reflections at the band edges 
accumulate to localize the electron or to create resonances; it is
far from being settled. Our contribution is to the question: how do
spectral properties change when
$\alpha$ is diminishing? We refer to \cite{ao} for a physical
discussion of this theme. Answers for two extreme cases are known: if
$\alpha>5/2$ then the spectrum of $H^{SW}$ is absolutely continuous
see \cite{6} and \cite{shab, brie} for generalizations; on the
other hand there are models (corresponding to $\alpha<0$) for which
the spectrum has no absolute continuous component,
\cite{nenc2,joye} or is even pure point \cite{aschduclexne}. If $V$
is analytic one has certain informations on existence and width of
resonances, \cite{grecsacc, aschbrie}.

The paper is organized as follows: in the next section we state our
main result, Theorem \ref{main}, precisely and infer Theorem
\ref{cont}. The third section contains the dynamical information:
the proof of Theorem \ref{main} organized in several subsections.

\section{Spectral properties}
Denote the free Stark Hamiltonian by $H^{SW}_0=-\Delta-x$; by
$D:=-i\partial_x$ the momentum operator which is 
selfadjoint on $H^1(\bR)$; $\chi$ is the binary fonction with
values $\chi(True)=1, \chi(False)=0$;
$\chi(D-t\in\lbrack a,b\rbrack)$ is the cutoff function in
Fourierspace of the interval
$\lbrack t+a,t+b\rbrack$;
$cte.$ a generic constant which may change from line to line;
\[\langle
n\rangle:=\left\lbrace\begin{array}{ll}1&n\le0\\
n&n>0\end{array}\right..\]

The main theorem is:
\begin{theorem}\label{main}
Let $\alpha>0, M>0$. There exists a $c=c_{\alpha,M}>0$ such that for
$V$ with $\Vert V\Vert_\alpha\le M$ and for all
$t\in\bR, n\in\bZ$ it holds:
\begin{itemize}
\item[(i)] \beqno&&\Vert\chi(D-t\in\lbrack n,
n+1))(e^{-iH^{SW}t}-e^{-iH^{SW}_0t})\Vert\le\\&& c_{\alpha,M}{\Vert
V\Vert_\alpha\over 
\langle sign(t) n\rangle^{min\lbrace1,\alpha\rbrace}};\eeqno

\item[(ii)] for $\psi\in L^2(\bR)$ this implies: 
\beqn{}&&\Vert\chi(D-t\in\lbrack n,
n+1))e^{-iH^{SW}t}\psi\Vert\ge\nonumber\\ &&\Vert\chi(D\in\lbrack
n,n+1))\psi\Vert- c_{\alpha,M}{\Vert V\Vert_\alpha\over 
\langle sign(t)
n\rangle^{min\lbrace1,\alpha\rbrace}}\Vert\psi\Vert.\label{HM}\eeqn
\end{itemize}
\end{theorem}

\begin{remark}
The dynamical meaning of (ii) is that a large part of a state whose
initial momentum is high enough is accelerated, i.e.: For any
$\varepsilon>0$ there is an
$n$ such that for
$\psi=\chi(D\in\lbrack n,n+1))\psi, \Vert\psi\Vert=1$ it holds
\[\Vert\chi(D-t\in\lbrack
n,n+1))e^{-iH^{SW}t}\psi\Vert\ge1-\varepsilon\qquad(t>0).\]
\end{remark}
We show now that the result on the spectrum follows from this:

\medskip
{\bf Proof of Theorem \ref{cont}.}
For a state $\psi$  in the pure point spectral subspace of
$H^{SW}$ it holds:
\[\lim_{R\to\infty}\sup_{\pm t\ge0}\Vert\chi(D\ge
R)e^{-iH^{SW}t}\psi\Vert=0,\]
see, for exemple \cite{enssvese}. This implies for $n\in\bZ$
\beq\label{BS}\lim_{t\to\pm\infty}\Vert\chi(D-t\in\lbrack
n,n+1))e^{-iH^{SW}t}\psi\Vert=0.\eeq
{\it ad (i):}  Consider  the initial state 
$\psi=\cF^{-1}\chi(p\in\lbrack n,n+1))$ where $\cF^{-1}$ denotes the
inverse unitary Fourier transform. By the inequality (\ref{HM}) it
holds for
$t>0,n$ large enough
\beqno\Vert\chi(D-t\in\lbrack
n,n+1))e^{-iH^{SW}t}\psi\Vert\ge1 -c{\Vert
V\Vert_\alpha\over
\langle sign(t)
n\rangle^{min\lbrace1,\alpha\rbrace}}\ge{1\over2}\eeqno  which
contradicts (\ref{BS}). So $\psi$ has a part in the continuous
subspace of $H^{SW}$.

\noindent{\it ad (ii):}
For $\psi$ in the pure point subspace take the limits $t\to sign (n)\
\infty$ 
 in  (\ref{HM}); the equality (\ref{BS}) implies
\[\Vert\chi(D\in\lbrack n,n+1))\psi\Vert\le c_{\alpha, M}{\Vert
V\Vert_\alpha\over\langle\vert
n\vert\rangle^{min\lbrace1,\alpha\rbrace}}\Vert\psi\Vert
\qquad(n\in\bZ)\]
which leads to
\[\Vert\psi\Vert^2=\sum_{n\in\bZ}\Vert\chi(D\in\lbrack n,
n+1))\psi\Vert^2\le c_{\alpha, M}\Vert
V\Vert_\alpha^2\sum_{n\in\bZ}{1\over\langle\vert
n\vert\rangle^{min\lbrace2,2\alpha\rbrace}}\Vert\psi\Vert^2\] which
is a contradiction for
$\Vert V\Vert_\alpha$ small enough and $\alpha>1/2$; so there are
no bound states. 
\hfill$\Box$

\section{Dynamics}
To prove Theorem \ref{main} we decompose the operator $H^{SW}$
in the Bloch representation.  Denote by
$\bV$ the convolution operator
$\bV\psi(n)={1\over\sqrt{2\pi}}\sum_{m\in\bZ}\widehat
V(n-m)\psi(m)$ in $L^2(\bZ)$.

$V$ is real, so it holds: $\overline{\widehat
V(n)}=\widehat V(-n)$. We can always substract a constant from 
$H^{SW}$ so we suppose that $\widehat V(0)=0$. 

Consider in 
\[L^2(\lbrack 0,1), dk; L^2(\bZ))\simeq\int_{\lbrack 0,1)}^\oplus
L^2(\bZ) \ dk\]
the time dependent operator
\[H(t)\psi(k,n)=(n+k+t)^2\psi(k,n)+\bV\psi(k,n).\]
$H_0(t)$ denotes this operator for $V=0$ which has constant domain
$H^2(\bZ)$.
$\bV$ is  bounded from $H^2(\bZ)$ to $L^2(\bZ)$ for $\alpha\ge0$
so the propagator  $\bU$ generated by $H(t)$ is well defined in
the strong sense. Its relation to the Wannier Stark propagator is:
\beq \bU(t)=\cB e^{-itx}e^{-iH^{SW}t}\cB^{-1}\label{BR}\eeq
where $\cB$ is the Fourier-Bloch transformation
\[\cB\psi(k,n)={1\over\sqrt{2\pi}}\int_0^{2\pi}e^{-inx}
\left\{\sum_{\gamma\in2\pi\bZ}
e^{-ik(x+\gamma)}\psi(x+\gamma)\right\}\ dx\]
which maps $L^2(\bR)$ unitarily onto $L^2(\lbrack 0,1), dk;
L^2(\bZ))$.

 Remark that $\cB\psi(k,n)=\cF\psi(k+n)$. We denote the
quantities for the case $V=0$  by a subscript $0$.

The following statement is a reformulation of Theorem \ref{main}
in this representation. 

Denote $P_n$ the projection on the the site $n$
in $L^2(\bZ)$ then: 
\begin{theorem}\label{tech}
For $\alpha>0, M>0$ there exists $c=c_{\alpha,M}>0$  such that for
$V$ with $\Vert V\Vert_\alpha\le M$ it holds in $\int_{\lbrack
0,1)}^\oplus L^2(\bZ) \ dk$ for $t\in\bR, n\in\bZ$:
\beqno {}&& \Vert P_n(\bU(t)-\bU_0(t))\Vert_\oplus=\\
&&\Vert\int_0^t P_n\bU^\ast_0\bV \bU\Vert_\oplus\le
c_{\alpha,M}{\Vert V\Vert_\alpha\over 
\langle sign(t) n\rangle^{min\lbrace1,\alpha\rbrace}}.\eeqno
\end{theorem}
We first prove that this is indeed equivalent to Theorem \ref{main}:

{\bf Proof of Theorem \ref{main}.}
By the identity
(\ref{BR}) it holds: 
\beqno{}&&\Vert\chi(D-t\in\lbrack
n,n+1))(e^{-iH^{SW}t}-e^{-iH^{SW}_0t})\Vert_{L^2(\bR)}=\\
{}&&\Vert\cB e^{-itx}\chi(D-t\in\lbrack
n,n+1))e^{itx}\cB^{-1}\cB e^{-itx}(e^{-iH^{SW}t}-e^{-iH^{SW}_0t})
\cB^{-1}\Vert_\oplus=\\
&&\Vert\chi(.+k\in\lbrack n,n+1))(\bU(t)-\bU_0(t))\Vert_\oplus=\\
&&\Vert P_n(\bU(t)-\bU_0(t))\Vert_\oplus.\eeqno
so part (i) is equivalent to Theorem \ref{tech}. To see part (ii)
observe that
\[e^{iH^{SW}_0t}\chi(D-t\in\lbrack
n,n+1))e^{-iH^{SW}_0t}=\chi(D\in\lbrack n,n+1))\]
so by (i) and the unitarity of $e^{iH^{SW}_0t}$:
\beqno{\ }&&\Vert\chi(D-t\in\lbrack n,
n+1))e^{-iH^{SW}t}\psi\Vert\ge\\
&&\Vert\chi(D-t\in\lbrack n,
n+1))e^{-iH^{SW}_0t}\psi\Vert- c_{\alpha,M}{\Vert
V\Vert_\alpha\over 
\langle sign(t)
n\rangle^{min\lbrace1,\alpha\rbrace}}\Vert\psi\Vert=\\
&&\Vert\chi(D\in\lbrack n, n+1))\psi\Vert- c_{\alpha,M}{\Vert
V\Vert_\alpha\over 
\langle sign(t)
n\rangle^{min\lbrace1,\alpha\rbrace}}\Vert\psi\Vert.\eeqno
 \hfill$\Box$

In the rest of the paper we prove Theorem \ref{tech}.

\subsection{\bf Proof of Theorem \ref{tech}}
 $\bU(t)=\int^\oplus U^k(t)\ dk$. For $\psi\in H^2(\bZ)$,
$\Omega^k:={U_0^k}^\ast U^k$ it holds
\beqno{}&&\Vert P_n(U^k(t)-U^k_0(t))\psi\Vert_{L^2(\bZ)}=
\Vert P_n(\Omega^k(t)-\bI)\psi\Vert\le\\
&&\Vert\int_0^tP_n {U_0^k}^\ast\bV U_0^k\Omega^k\psi\Vert.\eeqno
Remark that 
\[U^k(t,s)=U^0(t+{k},s+{k})\]
so it is sufficient to estimate
\beq\label{INT}\Vert\int_0^t P_n{U_0^0}^\ast\bV
U_0^0\Omega^0\psi\Vert.\eeq
We shall give the argument for nonnegative integers $n$ and
$t>0$. For $n<0, t<0$ the result then follows from
time reversal, i.e.: because for
$\bU(-t)=T\bU(t)T^{-1}$ with $T\psi(n):=\overline\psi(-n)$ it
holds:
$\Vert P_{-n}(\bU(-t)-\bU_0(-t))\Vert=\Vert
P_n(\bU(t)-\bU_0(t))\Vert$. For the remaining cases we shall give
an argument later.

In the sequel we shall drop the superscript $0$. We shall also
suppress the states $\psi$; the norm estimates below are to be
understood as uniform estimates  proven on the dense set
$H^2(\bZ)$ and valid by extension in the operator norm.

Rapid oscillations are responsible for the smallness of (\ref{INT}).
Remark firstly that with  
\[E_n(t):=(n+t)^2 \hbox{ it holds }\]
\[U_0(t)P_n=e^{-i\int_0^tE_n}P_n,\]
secondly that in every time interval
\[I_l:={l\over2}+\lbrack-{1\over4},{1\over4})
\qquad(l\in\bN)\] 
there is exactly one degeneracy, namely
\[E_n\left({l\over 2}\right)=E_{-n-l}\left({l\over 2}\right)\]
corresponding to some point of stationary phase of
$e^{-i\int_0^t(E_n-E_{-n-l})}$. Denote
\[\bP_{n,l}:=P_n+P_{-n-l}, \quad \bP_{n,l}^\bot:=\bI-\bP_{n,l}.\]
Heuristically, the contribution
\beq\label{TWOINT}
\Vert\int_{I_l}P_nU_0^\ast\bV U_0\Omega\Vert\le
\Vert\int_{I_l}P_nU_0^\ast\bV \bP_{n,l}U_0\Omega\Vert+
\Vert\int_{I_l}P_nU_0^\ast\bV \bP_{n,l}^\bot U_0\Omega\Vert\eeq
to (\ref{INT}) is small for two reasons; loosely speaking: the first
term describes the probability that a reflection from momentum $n$ to
$-n-l$ takes place, i.e. that the electron behaves adiabatically.
This are less likely for $n$ large and $\widehat V$ small.
The second term
describes transitions to the other states which is less probable
for $n$ large because the energetic distance to these states grows
like $\vert2n+l\vert$.

We shall now proceed to the proof according to this intuition. We
first treat the reflection to $-n-l$:
\begin{lemma}\label{cross}
There is a numerical constant $c>0$, independent of $V$, such that
for all $\alpha<\beta\in I_l$
\[\Vert\int_{\alpha}^\beta P_nU_0^\ast\bV U_0\bP_{n,l}\Omega\Vert\le
c\sqrt{1+\Vert V\Vert_0}{\vert\widehat V(2n+l)\vert\over
\sqrt{\ 2n+l}}\]
\end{lemma}

{\bf Proof.} We supposed that $\widehat V(0)=0$ so $P_n\bV P_n=0$;
with  the notation
\[\varphi(t):=\int_0^t(E_n-E_{-n-l})=\int_0^t(2n+l)(-l+2\tau)\
d\tau;
\quad\bV_{n,l}:=P_n\bV P_{-n-l};\]
we shall  estimate
\[\Vert\int_{I_l}e^{i\varphi }\bV_{n,l}\Omega\Vert.\]
It will be clear that the reasoning holds uniformly if we
integrate only on $\lbrack\alpha, \beta\rbrack\subset I_l$. This
estimate is done by a stationary phase calculation in the spirit of
\cite{bornfock}.

Decompose for an $a\in(0,1/2)$
\[I_l=I_l\setminus\left\lbrace{l\over2}+\lbrack-{a\over2},
{a\over2}\rbrack\right
\rbrace
\bigcup\left\lbrace{l\over2}+\lbrack-{a\over2},{a\over2}\rbrack
\right\rbrace =:I_l^{NS}\cup I_l^{S}.\]
Then as $\Vert\Omega\Vert=1$:
\[\Vert\int_{I_l^S}e^{i\varphi}\bV_{n,l}\Omega\Vert\le{a}\Vert\bV_{n,l}\Vert.\]
On the other hand an integration by parts of
$(\partial_t e^{i\varphi})\bV_{n,l}\Omega/i\dot\varphi$ yields
\beqno{}&&\left.\left\Vert\int_{I_l^{NS}}e^{i\varphi}\bV_{n,l}\Omega
\right\Vert\le
\left\Vert{\bV_{n,l}\over\dot\varphi}\right\Vert
\right\arrowvert_{\partial
I_l^{NS}}+\int_{I_l^{NS}}\left\Vert{\bV_{n,l}\dot\Omega\over\dot
\varphi}
\right\Vert+\left\Vert{\varphi^{\prime\prime}\bV_{n,l}\Omega\over\dot
\varphi^2}\right\Vert.\eeqno
Now observe that $\vert\dot\varphi(t)\vert=\vert 2n+l\vert\vert
-l+2t\vert\ge \vert 2n+l\vert a$, and that
$i\dot\Omega=U_0^\ast\bV U_0\Omega$, so the term on the last
expression is smaller than
\beqno{}&&4{\Vert \bV_{n,l}\Vert\over a\vert
2n+l\vert}+{\Vert \bV_{n,l}\Vert\over \vert
2n+l\vert}\int_{I_l^{NS}}\left({\Vert
P_{-n-l}\bV\Vert\over\vert
l-2t\vert}+{\vert
2\vert\over\vert
l-2t\vert^2}\right)\ dt\\
{}&&\le cte.{\Vert \bV_{n,l}\Vert\over a\vert
2n+l\vert}(1+\Vert V\Vert_0)\eeqno where we have used that
$\Vert P_{-n-l}\bV\Vert\le\Vert V\Vert_0$.
Thus for $a\in(0,1/2)$:
\[\left\Vert\int_{I_l}P_nU_0^\ast\bV U_0\bP_{n,l}\Omega\right\Vert\le
a\Vert\bV_{nl}\Vert+{1\over a}cte.{\Vert \bV_{n,l}\Vert\over \vert
2n+l\vert}(1+\Vert V\Vert_0)\]
The minimum of $a\alpha+\beta/a$ for positive $a$ is
$2\sqrt{\alpha\beta}$, thus
\[\left\Vert\int_{I_l}P_nU_0^\ast\bV\bP_{n,l}U_0\Omega\right\Vert\le
c{\Vert\bV_{n,l}\Vert\over
\sqrt{\vert 2n+l\vert}}\sqrt{1+\Vert V\Vert_0}\]
which implies our assertion as
$\Vert\bV_{n,l}\Vert\le{1\over\sqrt{2 \pi}}\vert
\widehat V(2n+l)\vert$.

\hfill$\Box$

\bigskip
The other levels are separated by large gaps. We start the proof
that the transition probability to them is small  with a double
integration by parts lemma.
 
We have $H_0(t)P_n=E_n(t)P_n$
 Denote by 
\[\widehat R_l(t):=(H_0(t)-E_n(t))^{-1}\bP_{n,l}^\bot\] 
the reduced resolvent. The Friedrichs
twiddle operation is fundamental in adiabatic theories. The version
needed here is defined for an operator $A$ on $L^2(\bZ)$ by
\[\widetilde{A}_l:=P_nA\widehat R_l.\]

\begin{lemma}\label{IBP}
For $\beta>\alpha$ it holds on $H^2(\bZ)$
\beqno{} &&\int_\alpha^\beta U_0^\ast P_n\bV \bP_{n,l}^\bot
U_0\Omega=\\ {}&&\int_\alpha^\beta
U_0^\ast\left({\widetilde{\widetilde
\bV_l \bV}}_l\bV+i\dot{\widetilde{\widetilde
\bV_l \bV}_l}-\widetilde\bV_l\bV
\bP_{n,l}-i\dot{\widetilde\bV}_l\right)U_0\Omega\\  {}&&
+\left.iU_0^\ast\left(\widetilde\bV_l-
\widetilde{\widetilde\bV_l\bV}_l\right)
U_0\Omega\right\arrowvert_\alpha^\beta\eeqno
\end{lemma}

{\bf Proof.} The twiddle operation is an inverse commutator.
A dot $\dot{}$ or a prime ${}^\prime$ denotes differentiation. It
holds:
\[i\partial_t U_0^\ast\widetilde{\bV}_lU_0=
U_0^\ast (\lbrack
\widetilde{\bV}_l, H_0-E_n\rbrack+i\dot{\widetilde{\bV}_l})U_0=
U_0^\ast( P_n\bV \bP_{n,l}^\bot+i\dot{\widetilde{\bV}_l})
U_0\]
so an integration by parts and $i\partial_t\Omega=U_0^\ast\bV
U_0\Omega$ yield
\beqno{} &&\int_\alpha^\beta U_0^\ast P_n\bV \bP_{n,l}^\bot
U_0\Omega=\\ {}&&-\int_\alpha^\beta
U_0^\ast\left({\widetilde\bV_l
}\bV+i\dot{\widetilde\bV}_l\right)U_0\Omega
+\left.iU_0^\ast\widetilde\bV_l
U_0\Omega\right\arrowvert_\alpha^\beta.\eeqno
The decomposition $\widetilde\bV_l\bV=\widetilde\bV_l\bV\bP_{n,l}+
\widetilde\bV_l\bV\bP_{n,l}^\bot$, the identity
\[i\partial_tU_0^\ast\widetilde{\widetilde\bV_l\bV}_lU_0=
U_0^\ast(\widetilde\bV_l\bV\bP_{n,l}^\bot+
i\dot{\widetilde{\widetilde\bV_l\bV}_l})U_0\]
and a second integration by parts imply
\beqno{} &&-\int_\alpha^\beta U_0^\ast
\widetilde{\widetilde\bV_l\bV}_lP_{n,l}^\bot U_0\Omega=\\
{}&&\int_\alpha^\beta
U_0^\ast\left(\widetilde{\widetilde\bV_l\bV}_l\bV+
i\dot{\widetilde{\widetilde\bV_l\bV}_l}\right)U_0\Omega
-\left.iU_0^\ast\widetilde{\widetilde\bV_l\bV}_l
U_0\Omega\right\arrowvert_\alpha^\beta.\eeqno
Thus the assertion is proved.
\hfill$\Box$

Concerning the second contribution to (\ref{TWOINT}) we shall now
proceed to estimate the different terms of 
\beq\label{off}\Vert\int_{I_l}P_nU_0^\ast\bV \bP_{n,l}^\bot
U_0\Omega\Vert\eeq
 defined by Lemma \ref{IBP} one after the other. 

In the following lemma we collect facts that shall be used
frequently and often without comment:

\begin{lemma}\label{estimates}
Let $a\neq b\in\bZ$.
For $\alpha, \beta>0$ there is a $cte.>0$ such that it holds
\[\sup_{j\in\bZ\setminus\lbrace a,
b\rbrace}{1\over \vert j-a\vert^\alpha\vert
j-b\vert^\beta}=\sup_{j<{a+b\over2}}\ldots+
\sup_{j\ge{a+b\over2}}\ldots\le cte. {1\over \vert
a-b\vert^{{min\lbrace \alpha, \beta\rbrace}}};\]
and for $\alpha, \beta>1$:
\[\sum_{j\in\bZ\setminus\lbrace a, b\rbrace}{1\over\vert
j-a\vert^\alpha}{1\over
\vert j-b\vert^\beta}\le\sum_{j<{a+b\over2}}\ldots+
\sum_{j\ge{a+b\over2}}\ldots\le
cte.{1\over\vert a-b\vert^{min\lbrace \alpha,\beta\rbrace}};\]
let $A$ be the operator on $L^2(\bZ)$ whose kernel is
$A(i,j)=f(i)g(i-j)$ for some $f,g\in L^2(\bZ)$, then
\[\Vert A\Vert=\sup_{\Vert
\varphi\Vert=\Vert\psi\Vert=1}\vert\langle\varphi,
A\psi\rangle\vert\le\Vert f\Vert_{L^2(\bZ)}\Vert
g\Vert_{L^2(\bZ)}.\]
\end{lemma}

The smallness of the terms in Lemma (\ref{IBP}) results from the
presence of the reduced resolvent; we shall use that for $m\neq n,
m\neq -n-l, t\in I_l$ it holds as
$\vert m+n+l\vert\ge1$ and so $\vert
m+n+l+\alpha\vert\ge{1\over2}\vert m+n+l\vert$ for
$\alpha<{1\over2}$:
\[\inf_{t\in I_l}\vert E_m(t)-E_n(t)\vert=
\inf_{I_l}\vert(m-n)(m+n+2t)\vert\ge{1\over 2}\vert
m-n\vert\vert m+n+l\vert\]
and as in  Lemma (\ref{estimates}):
\[\inf_{m\in\bZ\setminus\lbrace n, -n-l\rbrace}\inf_{t\in I_l}\vert
E_m(t)-E_n(t)\vert\ge cte.\vert n-(-n-l)\vert=cte.\vert
2n+l\vert.\]

The first relevant term in the integrand of Lemma (\ref{IBP}) is
$\widetilde{\widetilde\bV_l\bV}_l\bV=P_n\bV\widehat{R}_l\bV\widehat
R_l\bV$. Now
\[P_n\bV(i,j)=\delta_{ni}\widehat V(n-j);\quad\vert\widehat
R_l\bV(i,j)\vert=\chi(i\neq n)\chi(i\neq -n-l){\widehat V(i-j)\over
E_i-E_n}.\] 
By the third point of Lemma (\ref{estimates}) we get
\[\Vert\widehat R_l\bV\Vert\le\Vert
V\Vert_0\left(\sum_{\bZ\setminus\lbrace
n,-n-l\rbrace}\left\vert{2\over
(i-n)(i+n+l)}\right\vert^2\right)^{1/2}\le cte.\Vert
V\Vert_0{1\over\vert2n+l\vert}\]
so
\beq\label{A}\sup_{t\in I_l}
\Vert\widetilde{\widetilde\bV_l\bV}_l\bV\Vert\le\sup_{I_l}\Vert
P_n\bV\Vert\Vert\widehat R_l\bV\Vert^2\le cte.\Vert
V\Vert^3_0{1\over\vert 2n+l\vert^2}.\eeq
For the second term it holds
\[\Vert\dot{\widetilde{\widetilde\bV_l\bV}_l}\Vert=
\Vert(P_n\bV\widehat R_l\bV\widehat
R_l)^\prime\Vert\le2\Vert\bV\dot{\widehat R_l}\Vert\Vert\bV\widehat
R_l\Vert\]
so for $j\in\{n,-n-l\}$:
\[\sqrt{2\pi}\bV\dot{\widehat R_l}(i,j)={\widehat V(i-j)(\dot
E_j-\dot E_n)\over(E_j-E_n)^2}={2\widehat
V(i-j)\over(j-n)(j+n+2t)^2}\] so
\[\Vert\bV\dot{\widehat R_l}\Vert\le
cte.\Vert\bV\Vert_0{1\over\vert2n+l\vert}\]
and together with the estimate of $\bV\widehat R_l$ above it results:
\beq\label{B}\sup_{t\in
I_l}\Vert\dot{\widetilde{\widetilde\bV_l\bV}_l}\Vert\le cte.\Vert
V\Vert_0^2{1\over\vert2n+l\vert^2}.\eeq
 Next we discuss
$\dot{\widetilde\bV}_l=P_n\bV\dot{\widehat R_l}$.
\[\sqrt{2\pi}\dot{\widetilde\bV}_l(i,j)=2\delta_{in}{\widehat
V(i-j)\over(j-n)(j+n+2t)^2} {\vert i-j\vert^\alpha\over\vert
i-j\vert^\alpha}\] so:
\beq\label{C}\sup_{I_l}\Vert\dot{\widetilde\bV}_l\Vert\le
cte.\Vert V\Vert_\alpha{1\over\vert2n+l
\vert^{min{\lbrace1+\alpha,2\rbrace}}}.\eeq
The next contribution to (\ref{off}) comes from
\[\widetilde\bV_l\bV \bP_{n,l}=P_n\bV\widehat R_l\bV
P_n+P_n\bV\widehat R_l
\bV P_{-n-l}.\]
 The kernel of the
second term is
\[{2\pi}P_n\bV\widehat R_l\bV
P_{-n-l}(i,j)=\delta_{in}\delta_{j(-n-l)}
\sum_{\bZ\setminus\lbrace n,-n-l\rbrace}{\widehat V(n-m)\widehat
V(m+n+l)\over(m-n)(m+n+2t)}\]
so
\beqn&&\sup_{t\in I_l}\Vert P_n\bV\widehat R_l\bV
P_{-n-l}\Vert\nonumber\\ &&\le cte.
\sum_{\bZ\setminus\lbrace n,-n-l\rbrace}\left\vert{\widehat
V(n-m)\widehat V(m+n+l)\over(m-n)(m+n+l)}\right\vert{\vert
m-n\vert^\alpha\vert m+n+l\vert^\alpha\over\vert m-n\vert^\alpha\vert
m+n+l\vert^\alpha}\nonumber\\
&&\le cte.\Vert
V\Vert_\alpha^2{1\over\vert2n+l\vert^{1+\alpha}}\label{D}\eeqn
by the 
Cauchy-Schwartz inequality and Lemma (\ref{estimates}).

The other term is more difficult, it is the part which is first
scattered out of the state $n$, then back to it:
\[{2\pi}P_n\bV\widehat R_l\bV P_n(i,j)=
\delta_{in}\delta_{jn}\sum_{\bZ\setminus\lbrace
n,-n-l\rbrace}{\vert\widehat V(n-m)\vert^2\over(m-n)(m+n+2t)}.\]
Changing coordinates we obtain
\beqno &&{2\pi}P_n\bV\widehat R_l\bV P_n(i,j)=
\delta_{in}\delta_{jn}\sum_{p\in\bZ\setminus\lbrace
0,-(2n+l)\rbrace}{\vert\widehat V(p)\vert^2\over p(p+2n+2t)}=\\
&&{\vert\widehat V(2n+l)\vert^2\over
(2n+l)(4n+l+2t)}+\delta_{in}\delta_{jn}\sum_{p\in\bZ\setminus\lbrace
0,\pm(2n+l)\rbrace}{\vert\widehat V(p)\vert^2\over p(p+2n+2t)}=\\
&&{\vert\widehat V(2n+l)\vert^2\over
(2n+l)(4n+l+2t)}+\delta_{in}\delta_{jn}\sum_{p\in\bZ\setminus\lbrace
0,\pm(2n+l)\rbrace}-{\vert\widehat V(p)\vert^2\over p^2-(2n+2t)^2}
\eeqno
because only
$-{\vert\widehat V(p)\vert^2\over p^2-(2n+2t)^2}$,
the symmetric part of ${\vert\widehat V(p)\vert^2\over
p(p+2n+2t)}$, contributes to the symmetric sum. Physically speaking
it is destructive interference of the contributions of the
states $m=n\pm p$ which is at work here.

 Now for $\alpha>0$
\beqno&&\sup_{t\in I_l}\sum_{\bZ\setminus\lbrace
0,\pm(2n+l)\rbrace}{\vert\widehat V(p)\vert^2\over
\vert p^2-(2n+2t)^2\vert}{p^{2\alpha}\over p^{2\alpha}}\le\\
&&\Vert
V\Vert_\alpha^2\sup_{p,t}{1\over \vert p^{2\alpha}\vert\vert
p^2-(2n+2t)^2\vert}
\le\Vert V\Vert_\alpha^2
{cte.\over\vert2n+l\vert^{min\lbrace1+2\alpha,2\rbrace}};\eeqno
to see this recall that for $p\neq\pm(2n+l)$
\[\inf_{t\in I_l}\vert p^2-(2n+t)^2\vert\ge{1\over4}\vert
p^2-(2n+l)^2\vert,\]
and take the supremum over $\vert p\vert<{2n+l\over2}$ and $\pm
p\ge{\vert 2n+l\vert\over2}$ separately.

Furthermore
\[{\vert\widehat
V(2n+l)\vert^2\over\vert(2n+l)(2n+l+2n+2t)\vert}\le\Vert
V\Vert_0^2{cte.\over\vert2n+l\vert^2},\]
so the estimate for the backscattering term is
\beqn\label{E}\sup_{t\in I_l}\Vert P_n\bV\widehat
R_l\bV P_n\Vert&\le&cte. \sup_{t\in
I_l}\left\vert\sum_{\bZ\setminus\lbrace n,
-n-l\rbrace}{\vert\widehat
V(n-m)\vert^2\over(m-n)(m+n+2t)}\right\vert\nonumber\\
&\le&cte. \Vert
V\Vert_\alpha^2{1\over\vert2n+l\vert^
{min\lbrace1+2\alpha,2\rbrace}}.\eeqn
We are left with the boundary terms in (\ref{off}). We first discuss
\[\widetilde{\widetilde\bV_l\bV_l}
\left(t={l\over2}\pm{1\over4}\right).\]

$\widetilde{\widetilde\bV_l\bV_l}=P_n\bV_l\widehat
R_l\bV_l\widehat R_l$, so
\beq\label{F}\Vert\widetilde\bV_l\widetilde\bV_l({l\over2}\pm{1\over4})\Vert
\le\Vert\bV_l\widehat R_l({l\over2}\pm{1\over4})\Vert^2\le
cte.\Vert V\Vert_0^2{1\over\vert 2n+l\vert^2}\eeq
where we used the estimate which led to (\ref{A}).

For the other boundary term we have to be more careful:
Consider for $t_{l+1}:={l\over2}+{1\over4}$
\beqno&&
\sqrt{2\pi}\left(\widetilde\bV_l(t_{l+1})-\widetilde\bV_{l+1}(t_{l+1})\right)(i,j)=\\
&&P_n\bV(H-E_n)^{-1}(t_{l+1})(P_{-n-(l+1)}-P_{-n-l})(i,j)=\\
&&\delta_{in}\delta_{j(-n-l)}{\widehat
V(2n+l)\over{3/2}(2n+l+1)}-\delta_{j(-n-(l+1))}{\widehat
V(2n+l+1)\over{1/2}(2n+l)}.\eeqno
So
\beq\label{G}\Vert\widetilde\bV_l(t_{l+1})-\widetilde\bV_{l+1}(t_{l+1})\Vert\le
cte.\left({\vert\widehat
V(2n+l)\vert\over\vert2n+l\vert}+{\vert\widehat
V(2n+l+1)\vert\over\vert 2n+l+1\vert}\right).\eeq
Furthermore it holds for $x\in\lbrack-{1\over4},{1\over4})$:
\[\sqrt{2\pi}\widetilde\bV_l\left({l\over2}+x\right)(i,j)=\delta_{in}
{\widehat V(n-j)\over(j-n)(j+n+l+2x)}\chi(j\neq
n)\chi(j\neq-n-l),\]
so for $\alpha>0$
\beqn\label{H}\Vert\widetilde\bV_l({l\over2}+x)\Vert^2&\le&
cte.\sum_{\bZ\setminus\lbrace n, -n-l\rbrace}{\vert\widehat
V(n-j)\vert^2\over\vert(j-n)(j+n+l+2x)\vert^2}\nonumber\\
&\le&\Vert V\Vert^2_0
{cte.\over\vert2n+l\vert^{2}}.\eeqn

 With these observations we
have finished the proof of Theorem (\ref{main}). We assemble the
argument.

We have
\beqno&&\Vert\int_0^t P_nU_0^\ast\bV U\Vert_{L^2(\bZ)}=\\
&&\Vert\int_0^{1\over4} \dots\Vert+
\sum_{l=1}^{L-1}\Vert\int_{I_l} \dots\Vert+
\Vert\int_{{L\over2}-{1\over4}}^{t} \dots\Vert\le\\
&&\sum_{l=0}^\infty\Vert\int_{I_l} \dots\Vert\le\\
&&\sum_{l=0}^\infty\Vert\int_{I_l}
\dots\bP_{n,l}\Vert+\sum_{l=0}^\infty\Vert\int_{I_l}
\dots\bP_{n,l}^\bot\Vert\eeqno
where $L$ is the integer such that
$t\in\left\lbrack{L\over2}-{1\over4},{L\over2}+{1\over4}\right)$;
$I_0:=\left\lbrack0,{1\over4}\right)$; for the index $L$ we
have redefined
$I_L:=\left\lbrack{L\over2}-{1\over4},t\right)$. 

Remark that for $n>0, \beta>1$
\[\sum_{l=0}^\infty{1\over\langle2n+l\rangle^\beta}\le
{cte.\over\langle n\rangle^{\beta-1}}.\]

So by Lemma (\ref{cross})  it holds
\beqno&&\sum_{l=0}^\infty\Vert\int_{I_l}
\dots\bP_{n,l}\Vert\le cte.\sqrt{1+\Vert V\Vert_0}\sum
{\vert\widehat V(2n+l)\vert\over\sqrt{\langle 2n+l\rangle}}\\
&&\le cte.\sqrt{1+\Vert V\Vert_0}\Vert
V\Vert_\alpha\left(\sum{1\over\langle 2n+l\rangle^{2\alpha+1}}
\right)^{1/2}\\
&&\le cte.\Vert V\Vert_\alpha{1\over \langle n\rangle^\alpha}.\eeqno
By Lemma (\ref{IBP}) and the estimates (\ref{A}, \ref{B}, \ref{C},
\ref{D}, \ref{E}, \ref{F}, \ref{G}, \ref{H}) we have
\beqno&&\sum_{l=0}^\infty\Vert\int_{I_l}
\dots\bP_{n,l}^\bot\Vert\le\\
&&\sum \sup_{I_l}\left(\Vert{\widetilde{\widetilde
\bV_l \bV}}_l\bV\Vert+\Vert\dot{\widetilde{\widetilde
\bV_l \bV}_l}\Vert+\Vert\widetilde\bV_l\bV
\bP_{n,l}\Vert+\Vert\dot{\widetilde\bV}_l\Vert\right)+\\
&&\sum
\left(\Vert\widetilde{\widetilde\bV_l\bV}_l(t_{l+1})\Vert+\Vert
\widetilde{\widetilde\bV_l\bV}_l(t_l)\Vert\right)+\\
&&\sum\left(\Vert\widetilde \bV_{l}(t_{l+1})-\widetilde
\bV_{l+1}(t_{l+1})\Vert\right)+\Vert\widetilde
\bV_0(t_0)\Vert+\Vert\widetilde
\bV_L(t_L)\Vert\\
&&\le c_{\alpha, M}\Vert V\Vert_\alpha {1\over
\langle n\rangle^{min\lbrace1,\alpha\rbrace}}.
\eeqno
This proves the case $sign(n t)>0$. From our calculations it is
clear that for
$sign(n t)<0$ all the estimates give a bound proportional to
$\Vert V_\alpha\Vert$ and no decay in $n$. Thus the proof of
Theorem (\ref{main}) is finished.
\hfill$\Box$

\section{Acknowledgements}

This paper was partly written during the visits of G. Nenciu at CPT Marseille
and of F. Bentosela and P.Duclos at Univ. of Bucharest. Financial support
of CPT Marseille and the Romanian Ministry of Education (grant CNCSU 13C)
is hereby acknowledged.


\end{document}